\documentclass[12pt]{article}
\usepackage{epsfig}
\usepackage{epstopdf}
\usepackage{amsfonts}
\usepackage{amssymb}
\newcommand{\be}{\begin{equation}}
\newcommand{\ee}{\end{equation}}
\newcommand{\beq}{\begin{equation}}
\newcommand{\eeq}{\end{equation}}
\newcommand{\ben}{\begin{displaymath}}
\newcommand{\een}{\end{displaymath}}
\newcommand{\beqa}{\begin{eqnarray}}
\newcommand{\eeqa}{\end{eqnarray}}
\newcommand{\bea}{\begin{eqnarray}}
\newcommand{\eea}{\end{eqnarray}}
\newcommand{\bean}{\begin{eqnarray*}}
\newcommand{\eean}{\end{eqnarray*}}
\newcommand{\ba}{\begin{array}}
\newcommand{\ea}{\end{array}}
\newcommand{\bi}{\begin{itemize}}
\newcommand{\ei}{\end{itemize}}

\newcommand{\ie}{{\it i.e.,\,}}

\newcommand{\reef}[1]{(\ref{#1})}

\begin{document}

\begin{titlepage}
\begin{flushright}
hep-th/0702111 \\
MIT-CTP-3812
\end{flushright}
\vskip 1.5in
\begin{center}
{\bf\Large{Phases of Five-Dimensional Black Holes}}
 \vskip 0.5in {\bf Henriette Elvang$^{a}$, Roberto Emparan$^{b,c}$ and Pau Figueras$^{c}$}
 \vskip 0.3in
\textit{$^{a}$Center for Theoretical Physics}\\
\textit{Massachusetts Institute of Technology, Cambridge MA 02139, USA}\\
\medskip
\textit{$^{b}$Instituci\'o Catalana de Recerca i Estudis
Avan\c cats (ICREA)}\\
\medskip
\textit{$^{c}$Departament de F{\'\i}sica Fonamental}\\
\textit{Universitat de
Barcelona, Diagonal 647, E-08028 Barcelona,
Spain}

\vskip .3 in
{\tt elvang@lns.mit.edu, emparan@ub.edu, pfigueras@ffn.ub.es}
\end{center}
\vskip 0.5in

\baselineskip 16pt
\date{}

\begin{abstract} 

We argue that the configurations that approach maximal entropy in
five-dimen\-sional asymptotically flat vacuum gravity, for fixed mass
and angular momentum, are `black Saturns' with a central, close to
static, black hole and a very thin black ring around it. For any value
of the angular momentum, the upper bound on the entropy is equal to the
entropy of a static black hole of the same total mass. For fixed mass,
spin and area there are families of multi-ring solutions with an
arbitrarily large number of continuous parameters, so the total phase
space is infinite-dimensional. Somewhat surprisingly, the phases of
highest entropy are not in thermal equilibrium. Imposing thermodynamical
equilibrium drastically reduces the phase space to a finite, small
number of different phases.

\end{abstract} 

\end{titlepage} \vfill\eject

\setcounter{equation}{0}

\section{Introduction}

In this paper we study the phases of black holes in five-dimensional
gravity\footnote{More precisely: stationary vacuum solutions,
$R_{\mu\nu}=0$, which are asymptotically flat and regular on and outside
the black hole horizons.} with total mass $M$ and angular momentum $J$,
\ie phases in the microcanonical ensemble. A question of particular
interest is which configuration maximizes the entropy for given $M$ and
$J$.

In five dimensions, in addition to the Myers-Perry (MP) black holes with
horizon topology $S^3$ \cite{MP}, there exist black rings with topology
$S^1\times S^2$, which imply discrete non-uniqueness in a range of
parameters \cite{rings}. But black rings turn out to introduce a much
larger degeneracy through multi-black hole configurations. To understand
these, recall that the $S^1$-radius of a black ring can be made
arbitrarily large, for a given mass, by thinning the ring. So, given a
sufficiently thin and long ring, we can imagine putting a MP black hole
at its center. This increases slightly the centripetal pull on the
ring, but the effect can be counterbalanced by increasing its rotation.
These configurations, dubbed {\it black Saturns} have been explicitly
constructed and analyzed recently in \cite{ef}. The black ring can
actually encircle the central black hole quite closely, giving rise to
effects such as rotational dragging, but here we are more interested in
situations where the interactions between the ring and the central black
hole are small. In principle the construction method used in \cite{ef}
allows to systematically add an arbitrary number of rings, but the
complication grows enormously with each new ring. Solutions with two
black rings (without any central MP black hole) have been
constructed in \cite{diring}. 

A scatter-plot sampling of the parameter space of the exact solutions in \cite{ef}
showed regions of the phase diagram where black Saturns are the
entropically dominating solutions. We will argue here that this happens
throughout the entire phase diagram: for any values of $M>0$,
$J>0$, {\it the phase with highest entropy is a black Saturn}. 
Naively, one might have thought that a configuration with multiple black holes
should increase its entropy by merging them
all into a single black
hole. That this need not be so for a black Saturn follows from two simple
observations: 
\begin{enumerate}

\item Among all single black objects of a given mass $M$, the one with
maximal entropy is the static ($J=0$) spherical black hole.

\item A black ring of fixed mass can carry arbitrarily large spin
$J$ by making its $S^1$-radius large enough (and its $S^2$-radius small
enough). Conversely, there is always a thin black ring of arbitrarily
small mass with any prescribed value of $J$.

\end{enumerate} 
So we may say that static spherical black holes are the most
efficient black objects (since they use up a minimal mass) for carrying
entropy, and black rings are the most efficient ones for carrying spin.
So, given any values of $M>0$ and $J>0$, the total entropy will be
maximized by putting virtually all the mass in a central static black
hole, and having an extremely long, thin and light black ring,
carrying all the angular momentum. The total entropy of this black
Saturn configuration approaches asymptotically, in the limit of
infinitely thin ring, the entropy of a static black hole with the
same total mass. We shall argue that in fact for any given value of the
total angular momentum, there exist black
Saturns spanning the entire range
of areas
\beq
0<\mathcal{A}<\mathcal{A}_\mathrm{max}=\frac{32}{3}\sqrt{\frac{2\pi}{3}}(GM)^{3/2}\,.
\eeq
Here $\mathcal{A}_\mathrm{max}$ is the area of the static black hole.

We can also infer another remarkable feature of the five-dimensional
black hole phase space. Black Saturns with a single black ring exhibit
two-fold continuous non-uniqueness \cite{ef}: besides the total $M$ and
$J$, two other continuous parameters --- say, the mass of the black ring
and its dimensionless `thickness' parameter --- are needed in order to
fully specify the solution. Adding $n$ more rings introduces $2n$ more
continuous parameters. Moreover, a ring can have an effect as small as
desired on the total black Saturn. So for generic values of $M$, $J$ and
$\mathcal{A}\in (0,\mathcal{A}_\mathrm{max})$ there are black Saturns
with an arbitrarily large number of rings, and therefore characterized
by an arbitrarily large number of continuous parameters. 

The phase space thus shows a striking infinite intricacy. However, most
of these solutions, in particular the ones that maximize the total area,
are not in thermal equilibrium: the temperatures and angular velocities
need not be the same on disconnected components of the horizon. We find
that imposing thermodynamic equilibrium drastically reduces the phase
space to a finite number of families (perhaps only three), each specified
by a function $M(J,\mathcal{A})$.

\setcounter{equation}{0}

\section{Phasing in Saturn}
\label{sec:phasein}

\begin{figure}[t]
\centerline{\includegraphics[width=10cm]{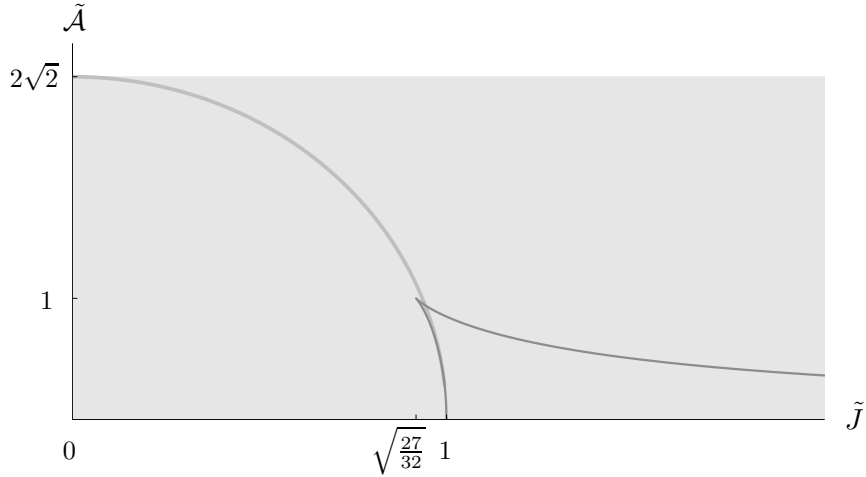}}
\begin{picture}(0,0)(0,0)
\put(30,58){\small $\tilde\mathcal{A}$}
\put(134,5){\small $\tilde J$}
\put(27,21){\footnotesize $1$}
\put(23,50.5){\footnotesize $2\sqrt{2}$}
\put(30,1){\footnotesize $0$}
\put(80,1){\footnotesize $1$}
\put(71,1){\footnotesize $\sqrt{\frac{27}{32}}$}
\end{picture}
\caption{\small Phases of five-dimensional black holes including
black Saturns. We fix total
mass $M=1$ and plot total area vs.\ spin. The solid curves correspond to
single MP black holes and black rings. The semi-infinite shaded strip,
spanning $0\leq \tilde J<\infty$, $0<\tilde\mathcal{A}<2\sqrt{2}$, is
covered by black Saturns. Each point in the strip actually corresponds
to a one-parameter family of Saturn solutions. The top end
$\tilde\mathcal{A}=2\sqrt{2}$ with $\tilde J\neq 0$ is reached only
asymptotically for black Saturns with infinitely long rings. Solutions
at the bottom $\tilde\mathcal{A}=0$ are naked singularities. For a fixed
value of $\tilde J$ we can move from the top of the strip to the bottom
by varying the spin of the central black hole $\tilde J_{h}$ from 0 to
1. For fixed area we can move horizontally by having $\tilde J_{h}<0$
and varying the spin of the ring between $|\tilde J_{h}|$ and $\infty$.}
\label{fig:phase}
\end{figure}

We begin by recalling the known single-black-object phases (\ie
with connected horizons), namely, MP black holes and black rings. 
We only consider solutions with angular momentum in a single rotation
plane.
In
order to eliminate awkward factors from the
formulas, we will work with rescaled spins and areas\footnote{For solutions
with a single black object of unit mass these coincide
with the reduced spin and area, $j$ and $a_{H}$, introduced in
\cite{re}.}
\beq \label{redef}
\tilde J\equiv\sqrt{\frac{27\pi}{32 G}}\,J\,,\qquad 
\tilde \mathcal{A}\equiv\sqrt{\frac{27}{256\pi G^3}}\,\mathcal{A}\,.
\eeq
The MP black hole phase with mass $M_h$ and angular momentum $J_h$ is 
characterized by the area
\beq\label{mp}
\tilde \mathcal{A}_{h}=2\sqrt{2\left(M_{h}^3-{\tilde J_{h}}^2\right)}\,,
\eeq
and the black rings with mass $M_r$, in parametric form, by area and 
angular momentum
\beq\label{bring}
\tilde \mathcal{A}_{r}=2\sqrt{M_{r}^3 \nu(1-\nu)}\,,\qquad 
{\tilde J_{r}}^2=M_{r}^3\frac{(1+\nu)^3}{8\nu}\,.
\eeq
To characterize their sizes, we introduce the circumferential radius of the
MP black hole in the rotation plane
\beq\label{rh}
R_{h}=2\sqrt{\frac{G}{3\pi}}\sqrt{\frac{2 M_{h}}{1-{\tilde J_{h}}^2/M_{h}^3}}
\eeq
and the $S^1$-radius of the inner rim of the ring,
\beq\label{r1}
R_1=2\sqrt{\frac{G}{3\pi}}\sqrt{M_{r}\frac{1-\nu}{\nu}}\,.
\eeq
The dimensionless parameter $\nu\in (0,1)$ is a `thickness' parameter
for the black ring.\footnote{For more precision and details see
\cite{eev}.} If we fix $M_{r}$, then as $\nu\to 0$ the angular momentum
and $R_1$ both go to infinity, and the area approaches zero: this is the
limit of an infinitely thin ring of infinite radius. In the opposite
limit, as $\nu\to 1$ we find a naked singularity with zero area. The
same singular solution is found in the extremal limit of MP black holes, $\tilde
J_{h}^2\to M_{h}^3$. The phase diagram with these solutions is shown in
figure \ref{fig:phase}. For every value of $\tilde J$, there is at least one black
object, and for $\sqrt{27/32}<\tilde J< 1$ there are three of them. 

Generically, adding spin for fixed mass reduces the area, as also
happens for the four-dimensional Kerr solution. So for a given value of $M$, the maximal
entropy is attained by the static black hole,
\beq\label{maxa}
\tilde\mathcal{A}_\mathrm{max}=\left(2M\right)^{3/2}\,.
\eeq

In principle any solution that is regular on and outside the horizon,
even if it consists of disconnected horizons, is an allowed phase. So
now we consider multi-black hole configurations. In classical General
Relativity we are free to fix a scale, which we will take to be the
total mass $M$. So, fixing 
\beq \label{mass1}
M=1 
\eeq 
we
want to know which solutions exist for a given total $J$, and what their
total entropy is. In particular, we want to find the
configurations with maximal entropy. 

The simplest black Saturns consist of a central MP black hole and a
single black ring. The actual solutions are rather complicated due to
the gravitational interactions between both objects, but if $R_1\gg R_h$
these effects become negligible. If $M_h$ is not much smaller than
$M_r$, this requires a very thin ring, \ie small $\nu$.\footnote{Thicker
rings with a tiny black hole at the center are of little interest to us
here, since they give phases with entropy very close to that of single
black rings.} Observe in particular that the dragging effect that the
black ring has on the central black hole can be made arbitrarily small,
no matter how large $\tilde J_{r}$, by making the ring thin and long
enough. This becomes clear by recalling that black rings resemble
boosted black strings, whose dragging falls off asymptotically in the
transverse radial direction $r$ like $\sim R_2/r$, where $R_2$ is the
radius of the $S^2$ of the string. So the effect near
the center of the ring is at most $\sim R_2/R_1\simeq \nu$. 

In this approximation, then, we can model a black Saturn as a simple
superposition 
of an MP black hole and a black ring. Take
\beq\label{toy}
\tilde\mathcal{A}=\tilde\mathcal{A}_{h}+\tilde\mathcal{A}_{r}\,,
\qquad M=1=M_{h}+M_{r}\,,\qquad  \tilde J=\tilde J_h+\tilde J_r\,.
\eeq
Comparing the plots of black Saturn phases produced using these
expressions with those from the exact solutions in \cite{ef} confirms
that the approximation is good when the rings are thin and long.

Having fixed the scale, configurations in this model have three free
parameters, just like in the actual black Saturn solutions. So for fixed
$(M,\tilde J)=(1,\tilde J)$, there is a two-parameter family of
solutions. Let us take $\tilde J_{h}$ and $\nu$ as these two parameters,
and ask which values of them
give maximal entropy. As discussed in the
introduction, we expect to achieve this by putting all the spin on a
thin, large ring, of very small mass. Graphical examination of the
function $\mathcal{A}(\tilde J_{h}, \nu; \tilde J)$ for different values
of $\tilde J$ shows indeed that the maximum area is achieved with
$\tilde J_h=0$ and $\nu \to 0$. So let us set $\tilde J_h=0$, and
consider small values of $\nu$. Then \beq\label{ajh0}
\tilde\mathcal{A}|_{\tilde J_h=0}= 2\sqrt{2}-6\sqrt{2}\,{\tilde
J}^{2/3}\nu^{1/3}+O(\nu^{2/3})\,. \eeq We indeed find that the maximal
area \reef{maxa} is reached from below as $\nu\to 0$ for {\it any} value
of $\tilde J$.

What is then the region of the plane $(\tilde J, \tilde\mathcal{A})$
covered when we include the Saturn phases? We now argue that it is the
semi-infinite strip
\beqa\label{strip}
&0<\tilde\mathcal{A}<\tilde\mathcal{A}_\mathrm{max}=2\sqrt{2}\,,&\nonumber\\
& 0\leq \tilde J<\infty\,,&
\eeqa
plus the point $(\tilde J, \tilde\mathcal{A})=(0,2\sqrt{2})$ for the
static black hole
(of course the symmetric region with $\tilde J<0$ is covered by reversing the
spins):

\begin{itemize}

\item We have seen that the upper boundary of the strip $(\tilde J>0,
\tilde\mathcal{A}=2\sqrt{2})$ is approached asymptotically as the
ring grows infinitely long and infinitesimally thin. 

\item The lower boundary $(\tilde J\geq 0, \tilde\mathcal{A}=0)$ corresponds
to naked singularities. One way to reach these is to make the central
black hole approach the extremal singular solution $\tilde J_{h}^2\to
M_{h}^3$, and adjusting the spin of the ring towards $\tilde
J-M_{h}^{3/2}$ while sending $M_r\to 0$ and $\nu\to 0$. Even if in this
limit $R_h\to\infty$, it is easy to see that, for any finite values of
the parameters, we can satisfy $R_1\gg R_h$ so our approximations hold.
Thus there are regular Saturn solutions arbitrarily close to the
lower boundary.

\item The solutions at the left boundary, with $(\tilde J=0,
0<\tilde\mathcal{A}\leq 2\sqrt{2})$ correspond to black Saturns where
the central black hole and the black ring are counterrotating. Such
configurations were studied in detail in \cite{ef}. To cover the entire
range $0<\tilde\mathcal{A}\leq 2\sqrt{2}$ we can simply have a central
black hole rotating to get the required area, and a thin black ring
counterrotating so that the angular momenta of the ring and the hole
cancel. It is possible to achieve $\tilde J=0$
also with fatter rings.

\item We can easily see that there is at least one black Saturn for any
point within these boundaries. Again, the idea is to have a central
black hole, in general spinning, accounting for the area
$\tilde\mathcal{A}$, and an extremely thin and long ring carrying the
required angular momentum to make up for the total $\tilde J$. 

\end{itemize}

So we can argue, by considering only black Saturns with thin rings, that
all of the strip \reef{strip} is covered. It is also easy to see why the
scatter-plots in \cite{ef} found it hard to sample the whole strip: the
solutions with the highest entropies are strongly localized in a small
region of the parameter space. Focusing and increasing the size of the
sample shows a clear tendency to cover the whole strip, and in
particular that entropies higher than those of rotating MP black holes
can occur for all $J\neq 0$.

In general there will be not just one black Saturn for every point
$(\tilde J, \tilde\mathcal{A})$ in the strip, but rather a one-parameter
family of them. These are guaranteed to exist by continuity, starting
from the configuration with an infinitely thin ring described above, and
continuously increasing $\nu$ while adequately adjusting $\tilde J_h$ to
keep $\tilde J$ and $\tilde\mathcal{A}$ fixed. This gives black Saturns
with fatter black rings, although at each point in the strip in general
there will be restrictions on the upper value for $\nu$, and on the
range of $\tilde J_h$.

When we discuss the configurations with maximal entropy, we should note
that our approximation based on \reef{toy} actually tends to
underestimate slightly the entropy of a black Saturn with a very thin
ring. Roughly, the reason is that the interaction between the central
black hole and the black ring is attractive, and so it decreases the
system's energy. Here we are fixing the total energy, which means that
the attraction leaves more rest-energy to be stored in the black hole,
giving a larger entropy than in the simple approximation \reef{toy}. We
have confirmed this effect with the exact solutions in \cite{ef}. It is
also possible to argue, by modeling the interaction in Newtonian terms,
that this does not allow to overshoot and reach areas larger than
$2\sqrt{2}$: for small $\nu$, the resulting increase in the area is at
least $O(\nu^{2/3})$ and hence smaller than the leading correction in
\reef{ajh0}. Other competing effects, like dragging, are even more
subdominant. So our conclusion that there do exist black Saturns with
area approaching arbitrarily close to a maximum \reef{maxa} seems robust
and holds even after including interactions.

\setcounter{equation}{0}

\section{Multiple rings and further parameters} 
\label{sec:multirings}

Using the same methods we can discuss solutions with multiple black
rings. Begin with two concentric black rings, and no central black hole.
The exact solutions have been built in \cite{diring}. When one of the
rings is much longer than the other we can use an approximation
analogous to \reef{toy} and simply superimpose two black rings.
Arguments of the kind above suggest that the maximal area will be that
of a minimally spinning ring (with $\nu=1/2$), \ie
$\tilde\mathcal{A}=1$. In this case, a strip of unit height, extending
to all $\tilde J\geq 0$, will be covered by these solutions. So some
double-ring configurations will have higher entropy than single rings
with the same mass and spin, with the maximal entropy achieved when the
innermost ring has $\nu\approx 1/2$ and the outer ring is very thin and
light. 

Black Saturns with multiple rings are clearly possible too, and when the
rings are thin and sparsely spaced we can model them
by adding more
terms to \reef{toy}. For solutions with many rings the cumulative
dragging effect may become stronger, but we can always have solutions
with thinner and thinner rings where this is negligible. 
Each class of solutions with fixed number $n$ of rings, and therefore
with $2n$ continuous parameters, covers the entire strip \reef{strip}. 

But there is even more. So far we have restricted ourselves to solutions
that rotate in a single plane. It is however possible to have both the
MP black hole and the black rings rotate in the plane orthogonal to the
ring plane. Black rings with two independent rotation parameters have
been constructed recently \cite{posen}. For fixed mass the $S^2$-spin
$J_2$ is bounded above, for reasons similar to the bound on Kerr black
holes. Furthermore, it can be seen that $|J_2|<|J_1|$. But another
consequence of having $J_2\neq 0$ is that $J_1$ is also bounded above,
and the extremal solutions that saturate this bound, for fixed mass,
have non-zero area. In other words, given $J_1, J_2\neq 0$ there is
a limit on how small the mass and how large the $S^1$ radius of the
black ring can be. But even if this second rotation puts an upper limit
on the $S^1$-radius of a black ring, it is clear that in general it adds
yet another set of continuous parameters to the phase diagram with fixed
$(M,J_1=J,J_2=0)$ that we have been studying. The second spin of the
central black hole, if not too large, can be cancelled against the
corresponding $S^2$-spin of a black ring. So with each black ring in the
black Saturn we get one more continuous parameter.

Still, multi-ring black Saturns may not exhaust the class of
asymptotically flat 5D vacuum black holes. Solutions that include
bubbles in combination with black holes \cite{weyl,eho} may play a role.
It has also been speculated that black holes with only one axial
symmetry might exist \cite{hr}. All of these, if they do exist, will add
further dimensions to the phase space, but we find it unlikely that they
can allow for larger total area than \reef{maxa}.

It is clear that the general phase space for solutions with non-zero
values of the two spins will, for the same reasons, be
infinite-dimensional. It will be interesting to analyze its overall
features.

\setcounter{equation}{0}

\section{First law of multi-black hole mechanics}

Consider a general stationary black Saturn (or any multi-black hole
solution) made of
$N$ black objects labelled by $i=1\,\dots,N$. Each connected component
of the horizon $H_i$ is generated by a Killing vector
\beq\label{kill}
k_{(i)}=\xi+\Omega_i \zeta\,,
\eeq
where $\xi$ and $\zeta$ are the canonically normalized Killing vectors
that generate time translations and rotations near infinity, and
$\Omega_i$ is the angular velocity on $H_i$. Following
standard procedure \cite{bch} we can write the ADM mass as a Komar
integral on a sphere
at infinity,
\beq
M=-\frac{3}{32\pi
G}\int_{S_\infty}
\epsilon_{abcde}\nabla^d\xi^e\,.
\eeq
Since we are in vacuum, Stokes' theorem allows us to write this as
\beq
M=-\frac{3}{32\pi
G}\sum_i\int_{H_i}
\epsilon_{abcde}\nabla^d\xi^e\,.
\eeq
Using \reef{kill}, a standard calculation leads to
the Smarr relation
\beq\label{smarr}
M=\frac{3}{2}\sum_i \left(\frac{\kappa_i}{8\pi G} \mathcal{A}_i +\Omega_i J_i\right)\,,
\eeq
where $\mathcal{A}_i$ and $\kappa_i$ are the area and
surface gravity on $H_i$, and
\beq
J_i=\frac{1}{16\pi G}\int_{H_i} \epsilon_{abcde}\nabla^d\zeta^e
\eeq
is the (Komar) angular momentum of the $i$-th black object.
Ref.~\cite{ef} checked that \reef{smarr} holds for the explicit solutions in
\cite{ef}. We could define
`Komar masses' for each black object
\beq
M_i=-\frac{3}{32\pi
G}\int_{H_i}
\epsilon_{abcde}\nabla^d\xi^e=\frac{3}{2}\left(\frac{\kappa_i}{8\pi G}
\mathcal{A}_i +\Omega_i J_i\right)\,, 
\eeq
but note that the latter equalities are actually mathematical identities
on each horizon, so they should not be thought of as `true' Smarr
relations.  

In a similar manner, a straightforward extension of \cite{bch} yields
the first law of multi-black hole mechanics\footnote{A first law and a
Smarr relation of this kind were found to hold in \cite{eho} for
explicit solutions with multiple static black holes in Kaluza-Klein
cylinders.}
\beq
\delta M=\sum_i\left( \frac{\kappa_i}{8\pi G}\; \delta\mathcal{A}_i
+\Omega_i\; \delta J_i\right)\,.
\eeq
The phase space of black Saturn configurations with up to $N$ black
objects is $2N$-dimensional. One may use the $(\mathcal{A}_i ,J_i)$ as
variables in it, or perhaps better, the total $\mathcal{A}$ and $J$, and
$(\mathcal{A}_i ,J_i)$, $i=2,\dots,N$. Fig.~\ref{fig:phase} is in fact a
projection of the phase diagram onto the plane $(\mathcal{A} ,J)$. It
must be noted that, as is apparent already when $N=1$, the
$(\mathcal{A}_i ,J_i)$ variables do not necessarily specify a unique
phase, as there may be discrete degeneracies.

\setcounter{equation}{0}

\section{Thermodynamical equilibrium}

Black Saturns are valid stationary solutions, regular on and outside the
horizons. But there is a glaring problem if they are to be considered as
thermodynamical phases: the separate black components have in general
different temperatures. Solutions with equal temperatures on all
horizons are possible, but in the configurations that maximize the
entropy the thin black ring has a temperature that is inversely
proportional to its $S^2$-radius and hence diverges in the limit of zero
thickness. So we find that the system can increase its entropy by
creating a large temperature gradient within itself! In other words, it
looks like the highest entropy state is not in thermal equilibrium,
which seems at odds with basic thermodynamical principles.

A similar problem arises with the horizon angular velocity $\Omega$,
which is the intensive `potential' for the angular momentum and
therefore would be expected to be homogeneous in a situation of
mechanical rotational equilibrium. For a very thin and light black
ring $\Omega\sim 1/R_1$ is very small, so in the black Saturns with
near-maximal area it is possible to arrange for $\Omega_h=\Omega_r$.
Nevertheless, generic stationary black Saturns have
a variety of different $\Omega_i$.

In order to understand better these discrepancies, recall that black
hole thermodynamics (as opposed to black hole mechanics) really makes
sense only when Hawking radiation is included. In the presence of
radiation, a generic black Saturn cannot be in thermal equilibrium, even
in the microcanonical ensemble. Even if $T$ is homogeneous, the
radiation cannot be in mechanical equilibrium between two black objects
with different values of $\Omega$. In fact the radiation will couple the
different black objects and drive the system towards thermodynamical
equilibrium. In the absence of this radiation each black object acts, to
some extent, as a separate thermodynamical system of its own. (A further
issue comes from the fact that black Saturns within wide parameter
ranges are expected to be classically dynamically unstable. We will
discuss further this point in the final section.)

\begin{figure}[t]
\centerline{\includegraphics[width=10cm]{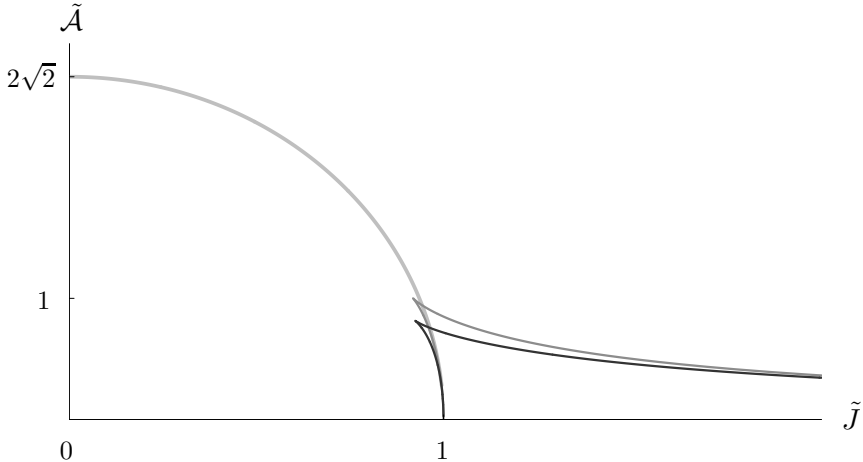}}
\begin{picture}(0,0)(0,0)
\put(30,58){\small $\tilde\mathcal{A}$}
\put(134,5){\small $\tilde J$}
\put(27,21){\footnotesize $1$}
\put(23,50.5){\footnotesize $2\sqrt{2}$}
\put(30,1){\footnotesize $0$}
\put(80,1){\footnotesize $1$}
\end{picture}
\caption{\small Phases in thermodynamical equilibrium. Besides the gray
curves for MP black holes and black rings, we have the solid black curve
of single-ring black Saturns for which the central black hole and the
black ring have equal temperature and equal angular velocity. There is
only one black Saturn (and not a continuous family of them) for each
point in this curve. The minimum $\tilde J$ along the curve is $\approx
0.9245$, slightly above $\sqrt{27/32}$, and the maximum total area is
$\tilde{\mathcal{A}} \approx 0.81$. Note that black Saturns are never
entropically dominant among the phases in thermodynamical equilibrium.
In the text we argue that it is unlikely that multi-ring black Saturns
exist as states in thermodynamical equilibrium. Barring more exotic
possibilities, this diagram may then contain all thermodynamical
equilibrium states of 5D black holes.}
\label{fig:phase2}
\end{figure}

If we consider a black Saturn with a single ring, then requiring equal
temperatures and angular velocities for the central black hole and the
black ring imposes two conditions on the parameters. This removes
entirely the continuous non-uniqueness, leaving at most discrete
degeneracies. Black Saturns in thermodynamical equilibrium thus form a
curve in the plane $(\tilde J, \tilde\mathcal{A})$. This curve,
generated from the exact solutions in \cite{ef}, is shown in
fig.~\ref{fig:phase2}.

The qualitative properties of the curve are easy to understand. Consider
a black Saturn with a thin (though not necessarily very thin) black
ring. In order for the temperature of the central black hole and the
black ring to be the same, the black hole radius must be roughly the
same as the ring's $S^2$ radius \cite{eev}. Then, the longer the ring,
the larger the fraction of the total mass it carries. In particular, as
$\tilde J\to \infty$ most of the mass is in the ring, so the black
Saturn curve asymptotes to that of a single black ring. As the black
ring gets thicker a larger fraction of the mass goes into the central
black hole, and the total entropy is only a fraction of the entropy of a
single ring. When the black ring is fat, the two black objects get
distorted and flattened along the rotation plane, approaching the
singular zero-area limit at $\tilde{J}=1$, common to black rings and MP
black holes. Although not shown in the figure, along the equilibrium
curve the angular velocity is smaller and the temperature higher than
those of a single ring of the same $\tilde{J}$. The reason is that in a
black Saturn the angular momentum is distributed over more objects
(hence $\Omega$ is lower) that are thinner (hence hotter) than in the
single black ring.

It looks much more difficult to impose equality of all temperatures and
all angular velocities on solutions with two or more black rings
(whether with a central MP black hole or not). The reason is that fixing
$T$ and $\Omega$ determines {\it uniquely} a black ring \cite{eev}. So
once we have one ring, if we try to put in another one with the same $T$
and $\Omega$, we are actually trying to put in {\it the same ring}.
There may be a slim chance that, due to non-linearities, a di-ring with
two very close-by rings of the same $T$ and $\Omega$ exist. This can be
settled by examining the exact solutions in \cite{diring}. But clearly
this is a strongly constrained possibility, and it seems highly unlikely
that an arbitrary number of black rings in thermodynamical equilibrium
can be piled on top of each other. 

So the condition of thermodynamical equilibrium not only removes all
continuous non-uniqueness, but it also reduces to a handful the families
of possible black objects: MP black holes, black rings, single-ring
black Saturns --- and this may be all. Barring the possibility of
multi-ring black Saturns, and of the conjectural solutions mentioned at
the end of section \ref{sec:multirings}, all the black hole phases in
thermodynamical equilibrium would be present in figure~\ref{fig:phase2}.
Up to a few degeneracies, these phases are specified by functions
$M(J,\mathcal{A})$. So the phase space in thermodynamic equilibrium is
two-dimensional, and shows only discrete non-uniqueness. 

Finally, note that when the second angular momentum is turned on, the
phase space of thermodynamic equilibrium becomes three-dimensional, with
energy function $M(J_1,J_2,\mathcal{A})$. It seems likely that this
phase space consists only of extensions of the families of solutions
considered above.

\setcounter{equation}{0}

\section{Discussion}

\begin{quote}
{\it For upon immediately directing my telescope at Saturn, I found that
things there had quite a different appearance from that which they had
previously been thought by most men to have.}
\begin{flushright}
{\sl Christiaan Huygens, Systema Saturnium} (1659)
\end{flushright}
\end{quote}
The phase structure we have uncovered has certainly quite a different
appearance from that which it had previously been thought to have.

An infinite-dimensional phase space, due to configurations with an
arbitrarily large number of black holes, is also known to occur in the
context of black holes localized in Kaluza-Klein cylinders, or `caged'
black holes \cite{kkbh}. However, configurations with multiple caged
black holes are all expected to increase their entropy by merging into a
single black hole. What is most remarkable about black Saturns is not so
much that they are possible, but that they maximize the entropy. The
essential idea behind this maximization consists of transferring all the
spin of the black hole (which tends to reduce the entropy) into orbital
angular momentum carried by a very light object at large radius. In
general this can be achieved through non-stationary configurations with,
say, faraway gravitational waves, or small black holes, carrying the
angular momentum, plus a central black hole close to maximal area. This
possibility was in fact considered in \cite{em}, but black Saturns
realize it solely in terms of stationary solutions.

Black Saturns with very thin and long rings are loosely bound
configurations, with negligible binding energy. But more importantly,
they are expected to be dynamically unstable. Very thin black rings are
likely to suffer from Gregory-Laflamme instabilities
\cite{gl,rings,hm,eev,hno}. If, as a result, the black ring fragments into
smaller MP-like black holes, these will fly apart. This final
configuration is reminiscent of those considered in \cite{em} for the
decay of ultra-spinning MP black holes in $D\geq 6$. There, the
possibility of increasing the area through fragmentation was one of
several telltale signs of a classical ultraspinning instability. We have
found that five-dimensional rotating MP black holes and black rings
might similarly increase their entropy through fragmentation into a
black Saturn (which may then fragment further). Could this signal a
classical instability of {\it all} five-dimensional black holes with
$J\neq 0$? 

Before jumping to this conclusion, it is important to bear in mind that
the existence of a configuration with higher total area is merely
suggestive of a possible dynamical instability. In addition, there must
exist some plausible classical dynamical evolution that drives the
initial configuration towards another one close enough to the putative
final state (through a curvature singularity, if fragmentation is to
occur). In this respect, it seems difficult for an MP black hole, or a
black ring, to dynamically evolve into a black Saturn of near-maximal
area, which has a black ring of {\it much larger} extent on the rotation
plane than the initial black hole. But it could possibly evolve towards
a black Saturn of roughly the same size as the initial black hole, if
the final area were higher. This remains to be investigated.

While it may seem striking that the configuration of maximal entropy is
classically unstable (except at $J=0$: the static black hole is stable
\cite{ik}), it should be recalled that it is not in thermal equilibrium,
so usual thermodynamic reasoning does not really apply here. When
Hawking radiation is taken into account, the very thin black ring of
black Saturn (or, if it fragments, the very small black holes flying
away) will evaporate much faster than the central black hole that
carries most of the mass. An important problem is to find the
dynamically stable configurations with maximal entropy among those that
are in stable thermodynamical equilibrium with Hawking radiation in the
microcanonical ensemble. It seems possible that this consists of an MP
black hole surrounded by spinning radiation. However, the thermodynamics
of other ensembles is likely to be quite different \cite{eev}. Another
question to bear in mind is that black Saturns with maximal area require
infinite volume. In a finite volume with fixed energy, maximal area may
not always be attained by a black Saturn. This will be the case when the
volume is large compared to, say, the Schwarzschild radius for that
energy. But if the box could barely fit a single MP black hole, or a
single black ring, then it might not fit a black Saturn of larger area.
All these issues must be sorted out depending on the specific application of
these ideas one is considering.

Dynamically stable black Saturns having only $M$ and $J$ as conserved
charges are likely to exist at least in Einstein-Maxwell theories. If
the black ring in the configuration carries a dipole charge \cite{re} it
can saturate an extremal (not BPS) bound, in which case the
Gregory-Laflamme instability is expected to disappear. Thin dipole black
rings have also been argued to be radially stable \cite{eev}. So dipole
black Saturns can presumably be made classically stable. Supersymmetric
multi-ring black Saturns, also expected stable, have been described in
\cite{gg}. It has been observed that in some cases they can have larger
entropy than a single black hole with the same conserved charges.
Semi-qualitative arguments similar to the ones in this paper might help
understand this effect.

Thin black rings have been argued to exist in any $D\geq 5$, achieving
equilibrium in essentially the same manner as in $D=5$ \cite{hm,eev}. So
black Saturns should exist, too, in any $D\geq 5$ and in particular the
ones that maximize the entropy should share the same features that we
have described in this paper. A semi-infinite strip in the $(\tilde J,
\tilde\mathcal{A})$ plane will then be covered by an infinite number of
families of solutions, but only a few of them
are expected to be in
thermal equilibrium.

Finally, our discussion of the phase space has been wholly classical. In
quantum gravity the continuous parameters will become discrete.
Furthermore, there will be a lower, Planck-length bound on the size of
the $S^2$ of thin black rings in a black Saturn. This will limit the
thinness of the ring and force the entropy of a black Saturn to be
finitely below the maximal value.

 \medskip
\section*{Acknowledgements}
\noindent
We thank Gary Horowitz for comments on the first version of this paper. HE was
supported by a Pappalardo Fellowship in Physics at MIT and by the 
US Department of Energy through cooperative research agreement 
DE-FG0205ER41360 . RE and PF were supported in part by DURSI 2005
SGR 00082, CICYT FPA 2004-04582-C02-02 and EC FP6
program MRTN-CT-2004-005104. PF was also supported by an FI scholarship
from Generalitat de Catalunya.


 \end{document}